\documentclass[conference]{IEEEtran}
\IEEEoverridecommandlockouts
\usepackage{cite}
\usepackage{amsmath,amssymb,amsfonts}
\usepackage{algorithmic}
\usepackage{graphicx}
\usepackage{textcomp}
\usepackage{xcolor}
\usepackage[hidelinks]{hyperref}
\hypersetup{
  colorlinks   = true, 
  urlcolor     = blue, 
  linkcolor    = black, 
  citecolor   = black 
}
\usepackage{multirow}
\usepackage{makecell}
\def\BibTeX{{\rm B\kern-.05em{\sc i\kern-.025em b}\kern-.08em
    T\kern-.1667em\lower.7ex\hbox{E}\kern-.125emX}}
\usepackage{setspace}
\usepackage{fancyhdr}
\usepackage{kantlipsum}
\fancypagestyle{plain}{
\fancyhf{}
\fancyhead[C]{Conference on \LaTeX} 
} \usepackage{eso-pic}

\makeatletter 
\def\@IEEEpubidpullup{8\baselineskip} 
\makeatother

\begin{document}

\IEEEoverridecommandlockouts
\IEEEpubid{
\parbox{\columnwidth}{\vspace{-4\baselineskip} Permission to make digital or hard copies of part or all of this work for personal or classroom use is granted without fee provided that copies are not made or distributed for profit or commercial advantage and that copies bear this notice and the full citation on the first page. Copyrights for third-party components of this work must be honored. For all other uses, contact the Owner/Author.\hfill\vspace{-0.8\baselineskip}\\ \begin{spacing}{1.2}
\small\textit{ASONAM '23}, November 6-9, 2023, Kusadasi, Turkey \\
\copyright\space2023 Copyright is held by the owner/author(s). \\
ACM ISBN 979-8-4007-0409-3/23/11 \\ \url{http://dx.doi.org/10.1145/3625007.3627477}
\end{spacing}
\hfill}
\hspace{0.9\columnsep}\makebox[\columnwidth]{\hfill}}
\IEEEpubidadjcol

\AddToShipoutPictureBG*{
\AtPageUpperLeft{
\setlength\unitlength{1in}
\hspace*{\dimexpr0.5\paperwidth\relax} 
\makebox(0,-0.75)[c]{\textbf{2023 IEEE/ACM International Conference on Advances in Social Networks Analysis and Mining (ASONAM)}}}}

\title{Measuring Online Emotional Reactions to Events}



\author{
Siyi Guo$^1$, Zihao He$^1$, Ashwin Rao$^1$, Eugene Jang$^2$, Yuanfeixue Nan$^2$, \\ Fred Morstatter$^1$, Jeffrey Brantingham$^3$, and Kristina Lerman$^1$\\
$^1$Information Sciences Institute,
$^2$University of Southern California,
$^3$University of California, Los Angeles\\
\{siyiguo, zihaoh, mohanrao,  eugeneja, ynan\}@usc.edu, fredmors@isi.edu, branting@ucla.edu, lerman@isi.edu
}


\maketitle

\begin{abstract}
The rich and dynamic information environment of social media provides researchers, policy makers, and entrepreneurs with opportunities to learn about social phenomena in a timely manner. However, using this data to understand social behavior is difficult due heterogeneity of topics and events discussed in the highly dynamic online information environment. To address these challenges, we present a method for systematically detecting and measuring emotional reactions to offline events using change point detection on the time series of collective affect, and further explaining these reactions using a transformer-based topic model. We demonstrate the utility of the method on a corpus of tweets from a large US metropolitan area between January and August, 2020, covering a period of great social change. We demonstrate that our method is able to disaggregate topics to measure population's emotional and moral reactions. This capability allows for better monitoring of population's reactions during crises using online data.
\end{abstract}


\section{Introduction}
Social media platforms connect billions of people worldwide, enabling them to exchange information and opinions, express emotions,  
and to respond to others. 
Researchers, policy makers, and entrepreneurs have grown interested in learning what the unfettered exchange of information reveals about current social conditions, 
including using social media data to track public opinion on  important issues~\cite{Barbera2015measuring} 
and monitor the well-being of populations at an unprecedented spatial scale and temporal resolution~\cite{pellert2022validating}. 

Using social media data to learn about human behavior, however, poses significant challenges. Social media represents a heterogeneous, highly dynamic information environment where some topics are widely discussed while others are barely mentioned \cite{dodds2022fame}. It includes people's self-reports of their own lives, as well as reactions to external events. Researchers have developed methods to 
detect events from online discussions, including clustering text based on similarity, analyzing term co-occurrence, identifying bursty terms and deep learning techniques \cite{malik2022performance,weng2011event,morabia-etal-2019-sedtwik,rezaei2022event}. 
However, social media data provides evidence for learning about human behavior beyond shifts in topics. For example, it can also shed light on emotions and morality, which are important drivers of individual attitudes, beliefs, and psychological and social well-being~\cite{vanKleef2016Editorial,haidt2007moral}.

To study the collective affect, researchers investigated how social media content influences emotional user engagement~\cite{babac2022emotion,aldous2022measuring}. 
These works, however, leave a gap in our understanding of collective emotional and moral reactions to socio-political events, which could shed light on opinion dynamics, emergence of polarization, and help identify online influence campaigns.
%

To bridge these gaps, we present a methodology for detecting, measuring and explaining the collective emotional reactions to offline events. Using state-of-the-art transformer-based models, we construct the time series of aggregate affect from social media posts. We detect emotional reactions as discontinuities in these time series, and then explain the reactions using topic modeling.
We demonstrate the utility of the methodology on a corpus of tweets collected from a large US metropolitan area between January and August, 2020. This time span represents a complex period in American history with important social, political and cultural changes. We successfully detect the simultaneous crises of the COVID-19 pandemic and racial justice reckoning, and other important events like political primaries. We show how these developments had profound impact on the psychological state of the population. For example, as the COVID-19 pandemic began to unfold, people expressed more {anger} and {fear}, and more moral sentiments like {care} and {authority}. Furthermore, we disaggregate COVID-related tweets by topic to more accurately measure the emotional reactions to different subtopics. We identify stronger reactions to daily-life issues (e.g. grocery panics) than topics directly mentioning COVID-19.

While we perform analyses on Twitter data, our pipeline is generalizable to other social media platforms and news. Our results suggest that studying the collective emotional reactions on social media can provide valuable insights into understanding people's opinions and responses to timely socio-political events, and aid policy makers in crafting messages that align with the values and concerns of the population\footnote{See the full version of this work \href{https://arxiv.org/abs/2307.10245}{here}.}.

\section{Methods and Materials}
\begin{figure}[h]
\centering
\includegraphics[width=1\linewidth]{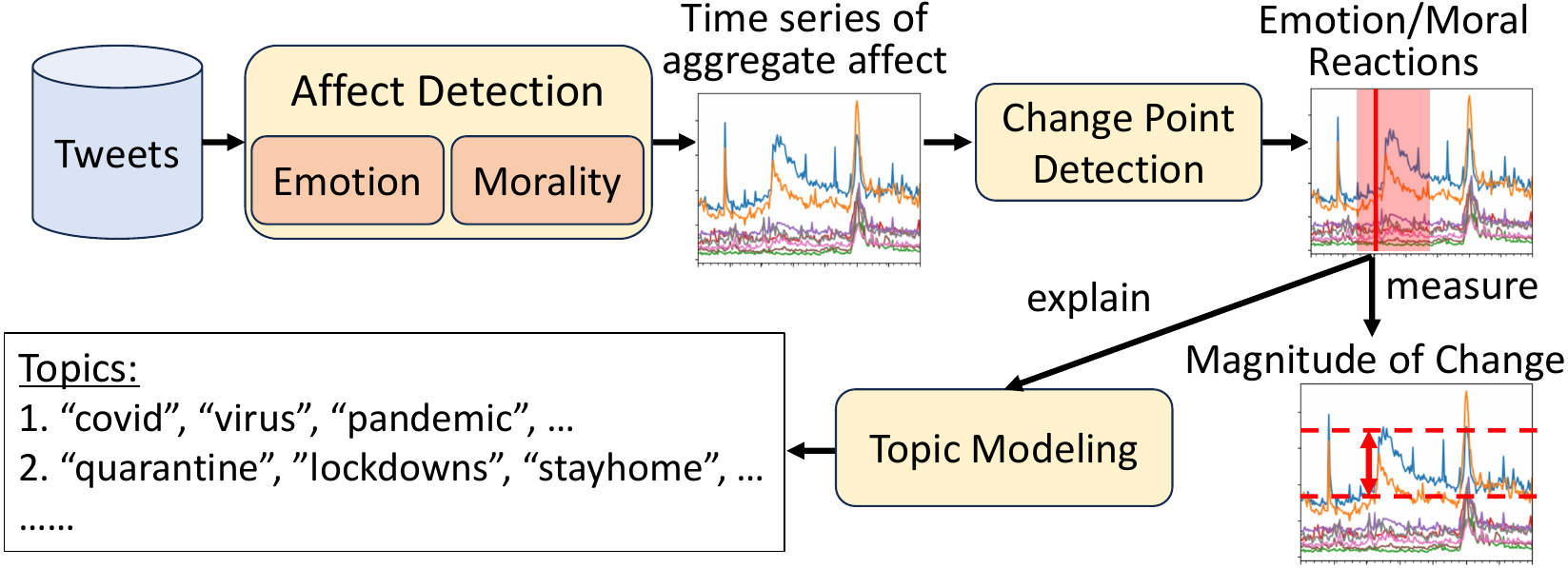} 
\caption{Pipeline to detect and measure online emotional reactions.}
\label{fig_flowchart}
\end{figure}

To understand the dynamics of affect, we propose a pipeline (Fig.~\ref{fig_flowchart}) that detects, measures and explains online emotional and moral reactions to offline events. With a set of timestamped texts, e.g. tweets, we first perform emotion and morality detection from text. We then 
construct the time series of the aggregate affect on a daily basis. Next, to detect reactions, 
we perform change point detection on each emotion and morality time series. We measure the magnitude of the change at each detected change point and perform topic modeling to explain the offline event that triggered the specific online reaction.

\subsection{Data}
The data used in this study was collected using Twitter's Filter API by specifying a geographic bounding box over 
a large metropolitan area. 
This method collects every tweet that is either geotagged within the bounding box (using the device's coordinates with the user's permission), or by using the Twitter ``place'' feature, where the user tags their location. 
We collected 17M tweets from 350K unique users.

\subsection{Emotion and Morality Detection}
We first measure emotions and moral sentiments expressed in an individual tweet. For emotion detection, we use a state-of-the-art language model SpanEmo~\cite{alhuzali-ananiadou-2021-spanemo}, fine-tuned on the SemEval 2018 1e-c data \cite{mohammad-etal-2018-semeval}. This transformer-based model outperforms prior methods by learning the correlations among the emotions. 
It measures 
\textit{anticipation}, \textit{joy}, \textit{love}, \textit{trust}, \textit{optimism}, \textit{anger}, \textit{disgust}, \textit{fear}, \textit{sadness}, \textit{pessimism} and \textit{surprise}.

We quantify the moral sentiments of tweets along five dimensions~\cite{haidt2007moral}: dislike of suffering (\textit{care}/\textit{harm}), dislike of cheating (\textit{fairness}/\textit{cheating}), group loyalty (\textit{loyalty}/\textit{betrayal}), respect of authority and tradition (\textit{authority}/\textit{subversion}), and concerns with purity and contamination (\textit{purity}/\textit{degradation}). 
We fine-tune a transformer-based model on diverse training data (see \cite{guo2023data} for details). 
The large amount and the variety of topics in our training data helps mitigate the data distribution shift during inference. After labeling tweets, we calculate the daily fractions of tweets with different emotion and moral categories to construct the time series.

We evaluate the effectiveness of emotion and morality detection on a random subset of 850 tweets, considering high difficulty and time cost of the annotation task. We asked five educated annotators to go through two training sessions, in each they annotated 50 random tweets and discussed to improve the agreement on the definitions of emotions and morality. Then each annotator individually annotated all 850 tweets. 
The Fleiss's $\kappa$ for emotion categories ranges in $0.42 \pm 0.02$. For morality categories, it ranges in $0.30 \pm 0.03$. Emotion and morality are very subjective concepts. Similar to prior works \cite{mohammad-etal-2018-semeval,Hoover2020moral}, we have found the $\kappa$ scores of some categories low. However, our agreement is still comparable and even better than these prior works. 

On this annotated dataset, we compare our emotion and morality detection methods with widely used dictionary-based methods, namely keyword matching using Emolex 
for emotions, and Distributed Dictionary Representations (DDR) \cite{8b2871f503a14011ae81e6ab1664a638} for morality. Our methods outperforms baselines on ten out of 11 emotion categories (the F1-scores of our method are in $0.42 \pm 0.14$, and those of Emolex are in $0.15 \pm 0.11$) and we outperform on nine out of ten moral categories (the F1-scores of our method are in $0.31 \pm 0.17$, and those of DDR are in $0.17 \pm 0.15$). 
The performance inevitably varies with support for different categories, as also observed in previous studies \cite{Hoover2020moral}. 
Despite the variation in model performance, prior research~\cite{pellert2022validating} has validated that when aggregating on the collective level, the time series of sentiments constructed with supervised deep learning detection and dictionary-based methods have strong correlations with those from self-reports.

\subsection{Detecting and Measuring Change Points} 
The time series of emotions and morality reveal the complex dynamics of aggregate affect on social media. We define an emotional reaction as a change in the corresponding time series. To detect such change points, we combine two popular methods. The first, cumulative sum (CUSUM) method~\cite{hinkley1971inference}, detects a shift of means, and is good at detecting changes like the COVID-19 outbreak, which shifted the baseline emotion and moral sentiment. To detect multiple change points, we use a sliding window to scan the whole time series. We set the window size to be four weeks and slide it every five days for the best precision. Another type of event, such as Valentine's Day, creates a short surge of emotions, can be better detected with Bayesian Online Change Point Detection (BOCPD)~\cite{adams2007bayesian}. It uses Bayesian inference to determine if the next data point is improbable, which is good at detecting sudden changes. We identify a change point to be significant when either CUSUM or BOCPD gives a significant confidence score, using $0.5$ as threshold. We perform change point detection separately for each time series of emotion and morality, because different types of events may elicit different reactions.  

For each detected change point, we quantify the magnitude of the collective reaction as percent change before and after it. We compute the baseline level before the change point as the mean of the time series over the two week period before. Then, we measure the short-term and long-term changes. To calculate the short-term change, we compare the baseline to the peak or dip value in the two weeks after the change point and compute percent change. To calculate the long-term change, we compare the baseline to the time series value two weeks after the event (we take a five-day average around the two-week mark). The size of the window is empirically chosen to be two weeks so that enough observations are made, but it would not be affected by another event earlier or later.

\subsection{Explaining Changes with Topic Modeling} \label{section_topic_modeling}

We try to explain changes in emotions detected by our method using topic modeling. 
We choose BERTopic~\cite{grootendorst2022bertopic}, a transformer-based language model that extracts highly coherent topics compared to traditional LDA. We evaluate both methods on a set of 10\% randomly selected tweets from our data, using a different numbers of topics ranging from 10 to 50 in steps of 10. Over different runs, BERTopic gives higher NPMI 
coherence scores ($0.14 \pm 0.01$) compared to LDA ($0.03 \pm 0.01$), and similar diversity \cite{dieng-etal-2020-topic} scores ($0.75 \pm 0.04$) compared to LDA ($0.76 \pm 0.04$).

For each emotional reaction, we extract the topics of tweets that are tagged with that emotion or morality category. 
We apply BERTopic to tweets within the three-day time window before and after the change, as discussions quickly die on social media~\cite{leskovec2009meme}. For example, for the Black Lives Matter (BLM) protests starting on 2020-05-26, we extract the topics from tweets posted between 05-23 to 05-25 to develop a baseline and then separately extract the topics between 05-26 to 05-28. By comparing the top 10 baseline topics before the change point with those after the change point, we determine the new topics that emerged after the change points that are possibly relevant to the event. 

For preprocessing, we remove URLs and name mentions, transform emojis to their textual descriptions, and split hashtags into individual words. We use the Sentence-BERT ``all-MiniLM-L6-v2'' model \cite{reimers-2019-sentence-bert} to directly embed the processed tweets. After topic modeling, we remove English stopwords in the learned topic keywords. 
With each emerging topic, we manually verify if there is an associated offline event by examining the tweets belonging to this topic and by searching related news articles. Such manual verification is a necessary and common practice event detection literature \cite{morabia-etal-2019-sedtwik}. 

\section{Results}

\subsection{Online Reactions to Offline Events}

\begin{figure}[!htbp]
\centering
\includegraphics[width=0.99\linewidth]{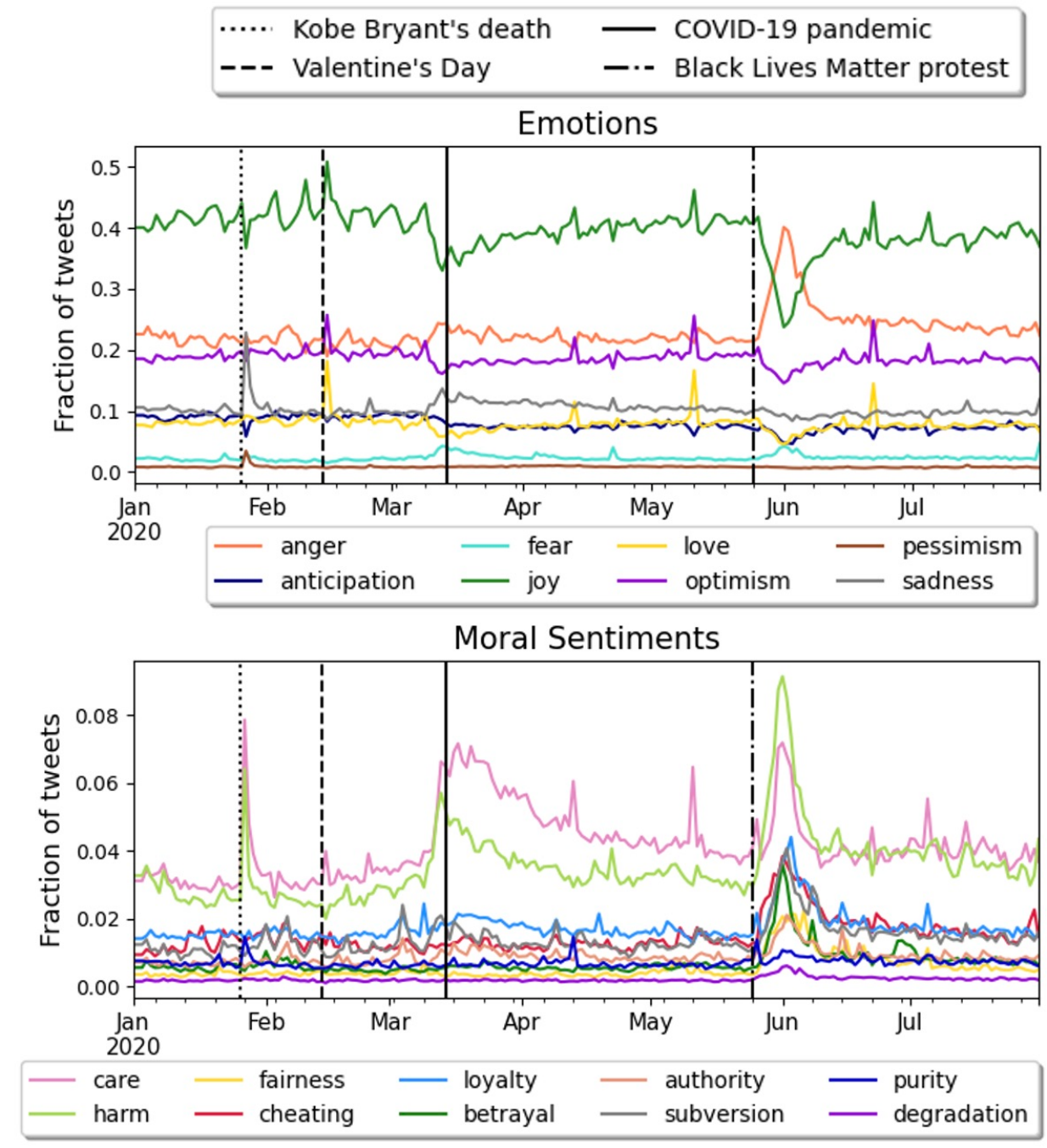} 
\caption{Time series of emotions and moral sentiments from January 1 to August 1, 2020. We show the daily fraction of tweets with different affect labels. The notable peaks and dips in the time series can be associated with the external events marked as vertical lines.}
\label{fig_ts}
\end{figure}

Time series of the aggregate affect from January to August 2020 (Fig.~\ref{fig_ts}) shows complex dynamics with seasonal variation (weekly cycles in joy), short-term bursts (spike in love on Valentine's Day), and long-term changes in emotions and moral sentiments. This time span represents a difficult period in the life of the city. In addition to the world-wide pandemic, which 
led to a national lockdown mid-March, political primaries were also taking place during this time period, which also saw one of the largest social justice protests triggered by the murder of George Floyd in police custody, as well as the death of a beloved sports icon. These developments had a profound impact on the city's population, as demonstrated by the many rises and dips in emotions and moral sentiments.

\begin{table*}[t]
\caption{Top events and the emotional and moral reactions detected by our method,  
in descending order of impact.}
\centering
\resizebox{1\linewidth}{!}{
\renewcommand{\arraystretch}{1.03}
\begin{tabular}{c|l|l|l|l|l|l}
\Xhline{1pt}
& \textbf{Event} & \textbf{Date} & \textbf{Time Window} & \textbf{Peaking Emotion/MF} & \textbf{Declining Emotion/MF} & \textbf{Relevant Topics} \\ \hline
1                     & \begin{tabular}[c]{@{}l@{}}MLK Jr. Day\end{tabular}                                             & 01-20                              & 01-17 to 01-20                                                                                  & joy, loyalty, fairness                                                                                                                                    &                                                                                                         & \begin{tabular}[c]{@{}l@{}}“mlk, king, martin”,\\ “rights, king, justice”\end{tabular}                                                \\ \hline
2                     & \begin{tabular}[c]{@{}l@{}}Trump’s impeachment\\ trial\end{tabular}                                              & 01-20                              & 01-20 to 01-21                                                                                  & betrayal, subversion                                                                                                                                      &                                                                                                         & \begin{tabular}[c]{@{}l@{}}“senate, vote, will”,\\ “treason, traitors, moscow”\end{tabular}                                           \\ \hline
3                    & Earthquake                                                                                                       & 01-20                              & 01-20                                                                                           & fear                                                                                                                                                      & \textbf{}                                                                                               & “earthquake, usgs, km”                                                                                                                \\ \hline
4                     & Kobe Bryant’s death                                                                                              & 01-26                              & 01-26                                                                                           & \begin{tabular}[c]{@{}l@{}}anger, pessimism, sadness,\\ care, harm, purity\end{tabular}                                                                   & joy                                                                                                     & \begin{tabular}[c]{@{}l@{}}“kobe, rest, peace”, \\ “prayers, praying, family”\end{tabular}                                            \\ \hline
5                     & \begin{tabular}[c]{@{}l@{}}Trump’s State of the\\ Union address\end{tabular}                                     & 02-03                              & 02-03                                                                                           & anger, disgust, authority                                                                                                                                 &                                                                                                         & \begin{tabular}[c]{@{}l@{}}“pelosi, nancy, speech”,\\ “pelosi, trump, tear”\end{tabular}                                              \\ \hline
6                     & Valentine’s Day                                                                                                  & 02-14                              & 02-14                                                                                           & love                                                                                                                                                      & anger, disgust                                                                                          & “valentine, valentines, happy”                                                                                                        \\ \hline
7                     & \begin{tabular}[c]{@{}l@{}}CA Pres. primary\end{tabular}                                       & 03-03                              & 03-02 to 03-03                                                                                  & fairness, subversion                                                                                                                                      &                                                                                                         & \begin{tabular}[c]{@{}l@{}}“vote, california”,“biden”\end{tabular}                                              \\ \hline
8                     & COVID-19 pandemic                                                                                                & 03-10                              & 03-09 to 03-11                                                                                  & \begin{tabular}[c]{@{}l@{}}anger, disgust, fear,\\ sadness, care, harm,\\ authority\end{tabular}                                                          & \begin{tabular}[c]{@{}l@{}}anticipation, joy,\\ love, optimism\end{tabular}                             & \begin{tabular}[c]{@{}l@{}}“coronavirus, corona, virus”,\\ “pandemic, virus, administration”,\\ “safe, stay, everyone”\end{tabular} \\ \hline
9                     & \begin{tabular}[c]{@{}l@{}}DOJ Drops Flynn Case\end{tabular}                       & 05-07                              & 05-07                                                                                           & cheating                                                                                                                                                  &                                                                                                         & “flynn, cheated, cheating”                                                                                                            \\ \hline
10                     & \begin{tabular}[c]{@{}l@{}}BLM protests\end{tabular}                                            & 05-26                              & 05-25 to 05-30                                                                                  & \begin{tabular}[c]{@{}l@{}}anger, disgust, fear, care,\\ harm, fairness, cheating,\\ loyalty, betrayal, authority,\\ subversion, degradation\end{tabular} & \begin{tabular}[c]{@{}l@{}}anticipation, joy,\\ love, optimism\end{tabular}                             & \begin{tabular}[c]{@{}l@{}}“racist, racism, white”,\\ “murder, george, floyd”, \\ “black, injustices, oppresion”\end{tabular}         \\ \hline

11                    & \begin{tabular}[c]{@{}l@{}}Dodgers beat Giants\end{tabular}                                              & 07-23                              & 07-23                                                                                           & joy                                                                                                                                                       & \textbf{}                                                                                               & “dodgers, mlb, giants”                                                                                                                \\ \hline

\Xhline{1pt}
\end{tabular}}

\label{tab_reactions}
\end{table*}

We ran the proposed pipeline to detect and explain the online emotional reactions to events. Table~\ref{tab_reactions} shows that our method is able to identify larger and impactful events such as the COVID-19 pandemic and the BLM protests. We see the complex reactions to the pandemic in multiple dimensions of emotions and morality. The unsupervised method also enables us to discover reactions to smaller events that might be easily missed, such as earthquakes and baseball playoffs. 
We also show that running BERTopic on tweets posted near the event reveals the relevant topics.
Further, because we detect changes separately in each emotion, we can disentangle events based on different emotional reactions, even when they take place on the same day: e.g., Trump's impeachment trial was associated with an increase in betrayal and subversion, MLK Day with joy, loyalty and fairness, and earthquake with fear.

\subsection{Evaluation of the Proposed Pipeline} \label{section_eval}
Similar to prior work~\cite{10.1145/2396761.2396785,morabia-etal-2019-sedtwik}, we use precision and duplicated event rate (DERate) to evaluate our method. Recall is not used because we cannot annotate every tweet to obtain an exhaustive list of events. Precision is the fraction of detected events that are related to realistic events~\cite{Allan1998}. We manually verify each event by searching news with topic keywords associated with each change point, which is common practice in event detection research. We detected 54 change points in total, with confidence  ranging from 0.65 to 1.00, and 85\% with confidence score above 0.9. We found 10 false positive, which cannot be explained by any topic and/or be related to any real event, giving a precision of 0.84.

Duplicate Event Rate (DERate) is the percentage of duplicate detected events among all realistic events detected~\cite{10.1145/2396761.2396785}. We define it as the fraction of emotion and morality categories (out of 21)  that detected the same event. The higher DERate shows better confidence. The DERate of our method is 0.14. Our precision and DERate are comparable to prior  works~\cite{morabia-etal-2019-sedtwik}.


\subsection{Short-term and Long-term Changes in Affect}

\begin{figure}[!t]
\centering
\includegraphics[width=0.884\linewidth]{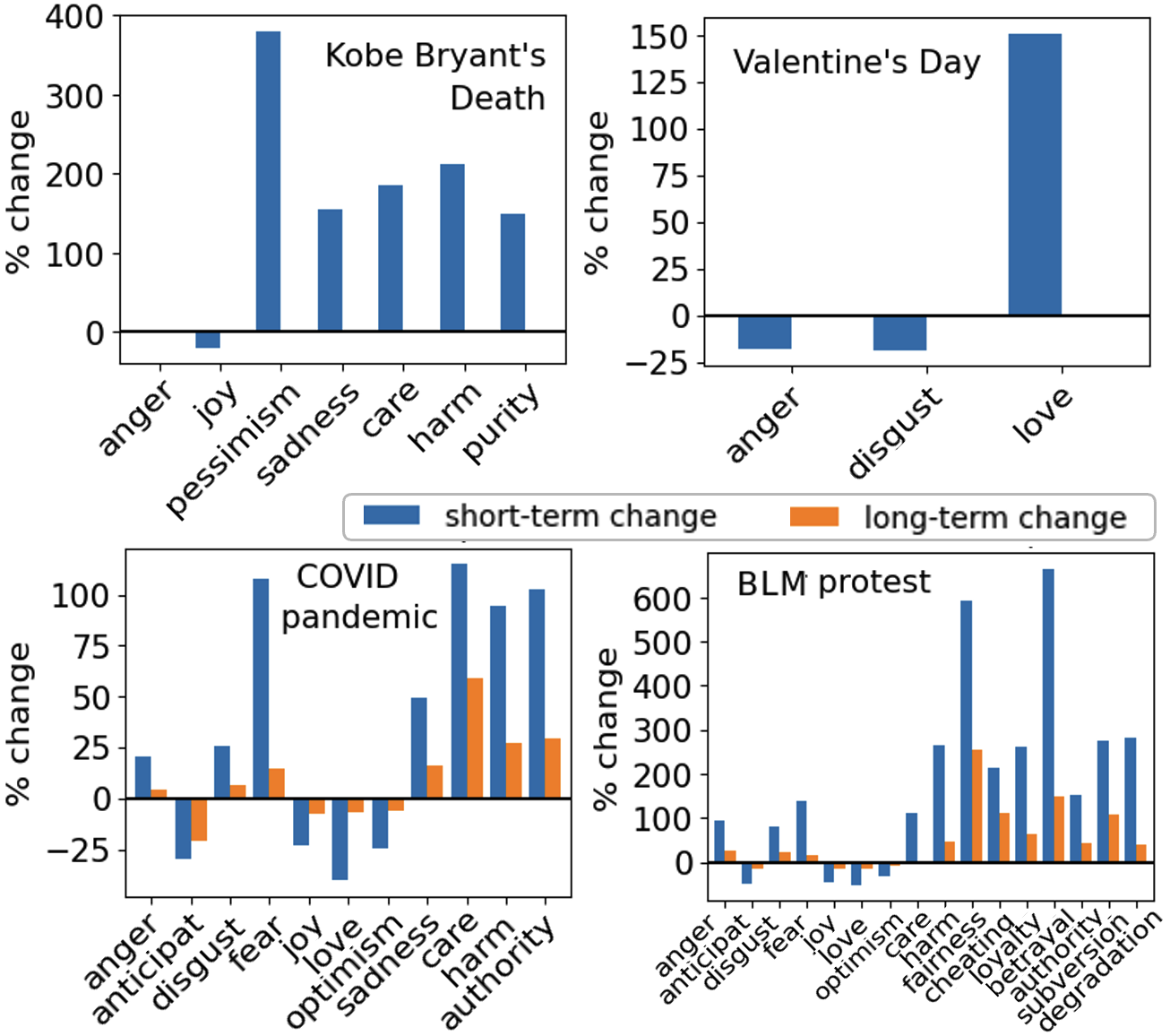} 
\caption{Short-term and long-term changes of emotions and moral sentiments around four events. The short-term change compares the peak/dip value after an event to the baseline level before the event. The long-term change compares the time series value around two weeks after the event to the baseline level.}
\label{fig_changes}
\end{figure}

Our proposed method enables us to study collective reactions to events along multiple dimensions of affect. For example, the BLM protests were associated with 16 different emotional and moral changes. We quantify the percent change in the corresponding collective affect before and after the event for four of the most impactful events (Fig.~\ref{fig_changes}). 
Consistent with our intuition, Kobe Bryant's Death was associated with a short-term increase in \textit{pessimism} and \textit{sadness} and a decrease in \textit{joy}, as well as a short-term rise in moral language related to \textit{care} and \textit{harm}. In contrast, Valentine's Day brought a short-term increase in \textit{love} and a decrease in \textit{anger} and \textit{disgust}. No long-term changes were seen with these events.

The COVID-19 outbreak triggered a cascade of events aimed at mitigating the pandemic that were associated with complex short-term and long-term changes in affect. People expressed more \textit{anger}, \textit{disgust}, \textit{sadness}, and more significantly, \textit{fear}, both in the short-term and the long-term. Positive emotions like \textit{joy} and \textit{love} simultaneously decreased. People also expressed more moral sentiments like \textit{care} such as in ``Stay safe. We thank you'', as well as more \textit{harm} blaming the virus. 
Interestingly, the moral language around \textit{authority} also increased, possibly due to new policies such as lockdowns to mitigate the pandemic (e.g. ``I think governor Newsom is doing a great job...''), and some were critical of government's response,  e.g., ``we need leadership not a politician''. 

The BLM protests was also associated with complex short- and long-term changes in affect. We observe increases in negative emotions and decreases in positive emotions. In addition, compared to other three events, we see greater increases in moral sentiments. The moral concerns about \textit{fairness} and \textit{betrayal} had especially increased, expressing a deep sense of the injustice and betrayal in George Floyd's death.


\subsection{Disentangling COVID-related Emotions}

\begin{figure}[h]
\centering
\includegraphics[width=0.92\linewidth]{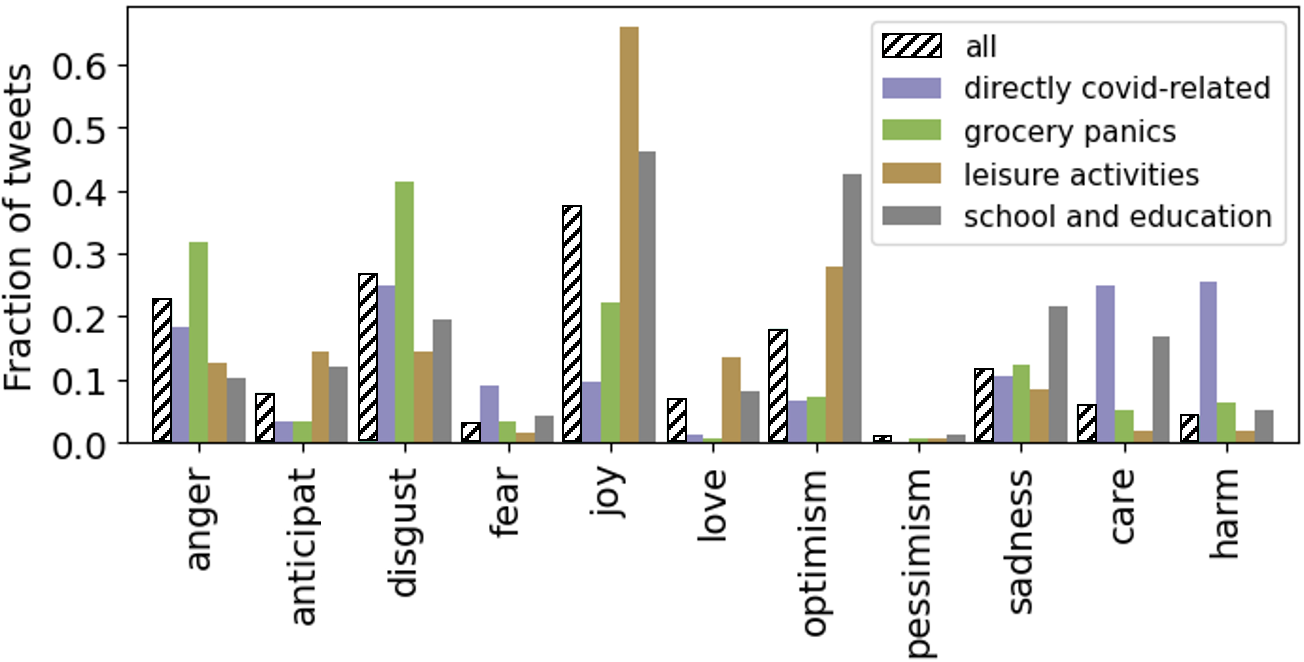} 
\caption{Emotions and moral sentiments expressed in COVID-related topics during the two weeks after WHO announcement of the pandemic on 2020-03-11. The topics are COVID (``coronavirus, corona, virus''); grocery panics (``grocery, groceries, shelves'', ``water, dasani, hydro'', and ``toilet, paper, rolls'');  leisure activities (``episode, episodes, show'', ``cook, cooking, cookout'',``tickets, ticket, selling''); and education (``teachers, students, learning'', ``schools, lausd, classes'', ``schools, lausd, closed''). 
}
\label{fig_covid_emot}
\end{figure}

The COVID-19 pandemic was associated with complex and long-term emotional changes. Here we use this example to further show the benefit of disentangling emotional reactions by disaggregating topics. We select four top categories discussed and group related topics into these categories: directly covid-related topics, grocery panics, leisure activities and school and education. 
We study emotions and moral expressions aggregated in all the tweets, as well as in these topic categories (Fig. \ref{fig_covid_emot}). 
We find that aggregating emotions from all tweets can give misleading impressions. Positive emotions like \textit{joy} were highly expressed in all tweets (aggregated), but in fact they were mostly dominated by people talking about leisure activities. In COVID-related tweets, few positive emotions were expressed. \textit{Anger} and \textit{disgust} were higher in topics about grocery panics than in topics directly related to COVID. Another example is the expression of \textit{care} and \textit{harm} moral sentiments. Their expressions were diluted by other topics in aggregate tweets. By disaggregating, we see that they were highly expressed in directly COVID-related tweets. These results suggest that during times of maximal crisis and uncertainty, people find outlets for positive emotions. They also demonstrate the importance of disaggregating by topics when studying specific issues.

\section{Conclusion}
In this work, we have demonstrated the effectiveness of an unsupervised method to detect and measure public reactions to newsworthy events. We applied our method to a large Twitter corpus of tweets drawn from the population of a large metropolitan area, disentangled the dynamics of online emotions during a time period punctuated by complex social, health, and political events. We showed that our method can discover significant and less significant events and measure emotional and moral reactions to these events. To further understand the complex impact of the COVID-19 outbreak, we disaggregated COVID-related tweets and discovered topics directly related to the virus and topics related to changes in life style, including unemployment, grocery panics and education. The emotions expressed on these different topics suggest that people had negative feelings and thoughts during the height of the pandemic, but were also searching for the positive  and holding on to optimism. Together, these results suggests the potential of using social media data fortracking of public reactions to events, as well as discovering significant events that may have been missed by traditional news sources.

There are some limitations of our method. When there is a change point that is a dip, we cannot use topic modeling to explain it, as a dip in the emotion or moral sentiment indicates a decrease of discussion related to an event. However, usually the decrease of some emotions is accompanied by the increase of some other emotions, and we can study the tweets tagged with the surged emotions to understand the topics.



\section*{Acknowledgment} This work is supported in part by AFOSR under grants FA9550-22-1-0380 \& FA9550-20-1-0224, and DARPA under contract HR001121C0168.

\bibliographystyle{IEEEtran}
\bibliography{main}

\begin{thebibliography}{10}
\providecommand{\url}[1]{#1}
\csname url@samestyle\endcsname
\providecommand{\newblock}{\relax}
\providecommand{\bibinfo}[2]{#2}
\providecommand{\BIBentrySTDinterwordspacing}{\spaceskip=0pt\relax}
\providecommand{\BIBentryALTinterwordstretchfactor}{4}
\providecommand{\BIBentryALTinterwordspacing}{\spaceskip=\fontdimen2\font plus
\BIBentryALTinterwordstretchfactor\fontdimen3\font minus \fontdimen4\font\relax}
\providecommand{\BIBforeignlanguage}[2]{{%
\expandafter\ifx\csname l@#1\endcsname\relax
\typeout{** WARNING: IEEEtran.bst: No hyphenation pattern has been}%
\typeout{** loaded for the language `#1'. Using the pattern for}%
\typeout{** the default language instead.}%
\else
\language=\csname l@#1\endcsname
\fi
#2}}
\providecommand{\BIBdecl}{\relax}
\BIBdecl

\bibitem{Barbera2015measuring}
M.~Klašnja \emph{et~al.}, ``{Measuring Public Opinion with Social Media Data}.''\hskip 1em plus 0.5em minus 0.4em\relax Oxford University Press, 09 2018.

\bibitem{pellert2022validating}
M.~Pellert \emph{et~al.}, ``Validating daily social media macroscopes of emotions,'' \emph{Scientific Reports}, vol.~12, no.~1, p. 11236, 2022.

\bibitem{dodds2022fame}
Dodds \emph{et~al.}, ``Fame and ultrafame: Measuring and comparing daily levels of ‘being talked about’ for united states’ presidents, their rivals, god, countries, and k-pop.'' \emph{JQD:DM}, vol.~2, Feb. 2022.

\bibitem{malik2022performance}
M.~Malik \emph{et~al.}, ``A performance comparison of unsupervised techniques for event detection from oscar tweets,'' \emph{Computational Intelligence and Neuroscience}, vol. 2022, 2022.

\bibitem{weng2011event}
J.~Weng \emph{et~al.}, ``Event detection in twitter,'' in \emph{In ICWSM-2011}, vol.~5, no.~1, 2011, pp. 401--408.

\bibitem{morabia-etal-2019-sedtwik}
K.~Morabia \emph{et~al.}, ``{SEDTW}ik: Segmentation-based event detection from tweets using {W}ikipedia,'' in \emph{NACCL workshop}, 2019, pp. 77--85.

\bibitem{rezaei2022event}
Z.~Rezaei \emph{et~al.}, ``Event detection in twitter by deep learning classification and multi label clustering virtual backbone formation,'' \emph{Evolutionary Intelligence}, pp. 1--15, 2022.

\bibitem{vanKleef2016Editorial}
G.~A. vanKleef \emph{et~al.}, ``Editorial: The social nature of emotions,'' \emph{Frontiers in Psychology}, vol.~7, p. 896, 2016.

\bibitem{haidt2007moral}
J.~Haidt \emph{et~al.}, ``The moral mind: How five sets of innate intuitions guide the development of many culture-specific virtues, and perhaps even modules,'' \emph{The innate mind}, vol.~3, pp. 367--391, 2007.

\bibitem{babac2022emotion}
M.~B. Babac, ``Emotion analysis of user reactions to online news,'' \emph{Information Discovery and Delivery}, no. ahead-of-print, 2022.

\bibitem{aldous2022measuring}
K.~K. Aldous \emph{et~al.}, ``Measuring 9 emotions of news posts from 8 news organizations across 4 social media platforms for 8 months,'' \emph{ACM Transactions on Social Computing (TSC)}, vol.~4, no.~4, pp. 1--31, 2022.

\bibitem{alhuzali-ananiadou-2021-spanemo}
H.~Alhuzali \emph{et~al.}, ``{S}pan{E}mo: Casting multi-label emotion classification as span-prediction,'' in \emph{ECACL}.\hskip 1em plus 0.5em minus 0.4em\relax ACL, Apr. 2021, pp. 1573--1584.

\bibitem{mohammad-etal-2018-semeval}
S.~Mohammad \emph{et~al.}, ``{S}em{E}val-2018 task 1: Affect in tweets,'' in \emph{Proc. 12th Int. Workshop on Semantic Evaluation}, Jun. 2018, pp. 1--17.

\bibitem{guo2023data}
S.~Guo \emph{et~al.}, ``A data fusion framework for multi-domain morality learning,'' in \emph{In ICWSM-2023}, vol.~17, 2023, pp. 281--291.

\bibitem{Hoover2020moral}
J.~Hoover \emph{et~al.}, ``Moral foundations twitter corpus: A collection of 35k tweets annotated for moral sentiment,'' \emph{Social Psychological and Personality Science}, vol.~11, no.~8, pp. 1057--1071, 2020.

\bibitem{8b2871f503a14011ae81e6ab1664a638}
J.~Garten \emph{et~al.}, ``\BIBforeignlanguage{English (US)}{Dictionaries and distributions: Combining expert knowledge and large scale textual data content analysis: Distributed dictionary representation},'' \emph{\BIBforeignlanguage{English (US)}{Behavior Research Methods, Instruments, and Computers}}, vol.~50, no.~1, pp. 344--361, Feb. 2018.

\bibitem{hinkley1971inference}
D.~V. Hinkley, ``Inference about the change-point from cumulative sum tests,'' \emph{Biometrika}, vol.~58, no.~3, pp. 509--523, 1971.

\bibitem{adams2007bayesian}
R.~P. Adams \emph{et~al.}, ``Bayesian online changepoint detection,'' \emph{arXiv preprint arXiv:0710.3742}, 2007.

\bibitem{grootendorst2022bertopic}
M.~Grootendorst, ``Bertopic: Neural topic modeling with a class-based tf-idf procedure,'' \emph{arXiv preprint arXiv:2203.05794}, 2022.

\bibitem{dieng-etal-2020-topic}
A.~B. Dieng \emph{et~al.}, ``Topic modeling in embedding spaces,'' \emph{TACL}, vol.~8, pp. 439--453, 2020.

\bibitem{leskovec2009meme}
J.~Leskovec \emph{et~al.}, ``Meme-tracking and the dynamics of the news cycle,'' in \emph{Proceedings of the 15th ACM SIGKDD}, 2009, pp. 497--506.

\bibitem{reimers-2019-sentence-bert}
N.~Reimers \emph{et~al.}, ``Sentence-bert: Sentence embeddings using siamese bert-networks,'' in \emph{In EMNLP-2019}.\hskip 1em plus 0.5em minus 0.4em\relax ACM, 11 2019.

\bibitem{10.1145/2396761.2396785}
C.~Li \emph{et~al.}, ``Twevent: Segment-based event detection from tweets,'' ser. CIKM '12.\hskip 1em plus 0.5em minus 0.4em\relax New York, NY, USA: ACM, 2012, p. 155–164.

\bibitem{Allan1998}
J.~Allan \emph{et~al.}, ``{Topic Detection and Tracking Pilot Study Final Report},'' 1998.

\end{thebibliography}

\end{document}